\documentclass[journal]{IEEEtran}

\usepackage{url}
\usepackage{hyperref}
\usepackage{multirow}
\usepackage{subfig}
\usepackage{lipsum}
\usepackage{verbatim}
\usepackage{graphicx}
\usepackage{epstopdf}
\usepackage{amsmath}
\usepackage{amsfonts}
\usepackage{balance}
\usepackage{tabularx}
\usepackage{verbatim} 
\usepackage[export]{adjustbox}

\usepackage{tikz}
\usetikzlibrary{positioning}
\ifCLASSINFOpdf
\else
\fi
\begin{document}
%
\title{ DNN-based cross-lingual voice conversion using Bottleneck Features}
%
%
%

\author{M~Kiran~Reddy~and~K~Sreenivasa~Rao
        
\thanks{The authors are with the Department of Computer Science and Engineering, Indian Institute of Technology, Kharagpur, West Bengal, India (email: kiran.reddy889@gmail.com; ksrao@iitkgp.ac.in).}
\thanks{}}
%
%

\markboth{}%
{Shell \MakeLowercase{\textit{et al.}}: Bare Demo of IEEEtran.cls for IEEE Journals}
%



\maketitle

\begin{abstract}
Cross-lingual voice conversion (CLVC) is a quite challenging task since the source and target speakers speak different languages. This paper proposes a CLVC framework based on bottleneck features and deep neural network (DNN). In the proposed method, the bottleneck features extracted from a deep auto-encoder (DAE) are used to represent speaker-independent features of speech signals from different languages. A DNN model is trained to learn the mapping between bottleneck features and the corresponding spectral features of the target speaker. The proposed method can capture speaker-specific characteristics of a target speaker, and hence requires no speech data from source speaker during training. The performance of the proposed method is evaluated using data from three Indian languages: Telugu, Tamil and Malayalam. The experimental results show that the proposed method outperforms the baseline Gaussian mixture model (GMM)-based CLVC approach. 
\end{abstract}


%
\IEEEpeerreviewmaketitle

\section{Introduction}

\IEEEPARstart{V}{oice} conversion (VC) is the process of modifying the speech utterances of a source speaker so that it sounds like it was uttered by a target speaker. There are
several applications of VC such as voice restoration, customization of text-to-speech systems, etc~\cite{b1}. Based on the language that source and target speakers speak VC can be divided into two categories, namely, (1) \textit{Intra-lingual VC} and (2) \textit{cross-lingual VC}. Intra-lingual VC assumes that the source and target speakers speak in the same language. Several techniques have been proposed in the literature for intra-lingual VC based on Gaussian mixture models (GMMs)~\cite{b1,b2,b3}, deep neural networks (DNNs)~\cite{b4,b5,b6}, recurrent neural networks (RNNs)~\cite{b7}, restricted Boltzmann machines (RBMs)~\cite{b8,b9}, exemplar-based framework~\cite{b10} and generative adversarial nets (GANs)~\cite{b11,b12,b13}. These approaches either require parallel or non-parallel data to obtain the mapping function between source and target spectral features. 

Cross-lingual VC (CLVC) is essential for developing mixed-language speech synthesis systems, customization of speaking devices, etc~\cite{b18}. In CLVC, it is assumed that the source and target speakers speak different languages (the languages should be acoustically closer), and the aim is to convert the utterance spoken by source speaker in such a way that it is spoken untranslated by the target speaker. Hence, there is no possibility of recording parallel data. Since, the languages spoken by source and target speakers are different, it is difficult to train conversion model even using non-parallel data. In the literature, very few attempts have been made to develop CLVC systems. In~\cite{b14}, vector quantization (VQ) based CLVC is developed for the languages, Japanese and English. This approach do not sufficiently preserve the speaker’s identity, where the feature space of the converted envelope is limited to a discrete set of envelopes. In~\cite{b15}, vocal tract length normalization is empolyed for developing CLVC system for the languages German and English. Erro et al., employed an iterative frame selection approach to perform cross-language VC~\cite{b16}. An eigenvoice (EV) GMM-based method for CLVC is proposed in~\cite{b17}. Here, parallel data from source speaker and multiple pre-stored data from other speakers are used for training EV-GMM. The EV-GMM is adapted using few arbitrary utterances from the target speaker in a different language, to obtain the conversion model. 

Though these methods avoid the need for parallel data, they still require non-parallel speech data from the source speakers to build the conversion models. This is a limitation to an application where an arbitrary source speaker's voice has to be transformed to a target speaker without recording anything \textit{apriori}. Thus, recent research focus on investigating conversion models which can capture target speaker-specific characteristics, and avoid the need for source speaker's data in training stage. In~\cite{b18}, speaker-specific GMMs are trained using data from target speaker alone, to achieve CLVC for Indian languages. This approach can convert speech of an arbitrary source speaker into a given target speaker. However, it suffers from the oversmoothing effect due to GMM-based based conversion. Recently, attempts are being made to capture speaker-specific characteristics using neural network (NN) models. Training speaker-specific NN models is not as straightforward as in the case of GMM. Here, first we need to efficiently extract both speaker-dependent and speaker-independent features from speech signals~\cite{b4}. Then, a NN can be used to obtain the mapping function between the speaker-dependent and speaker-independent features  computed from speech signals of the target speaker. While Mel-cepstral coefficients (MCEPs) can be used as speaker-dependent features, there is search for an efficient representation of speaker-independent information. Phonetic posteriorGrams (PPGs) have been explored as speaker-independent features for intra-lingual~\cite{b19} and cross-lingual VC~\cite{b20}. PPGs are estimated by training an ASR system using a large multi-speaker database. Although PPGs perform well for intra-lingual VC, their extraction become difficult in a cross-lingual scenario~\cite{b20}. Hence, there is a need for an alternative representation for speaker-independent information. 

\begin{figure*}[]
\centering
	\subfloat[]{%
  		\includegraphics[scale = 0.28]{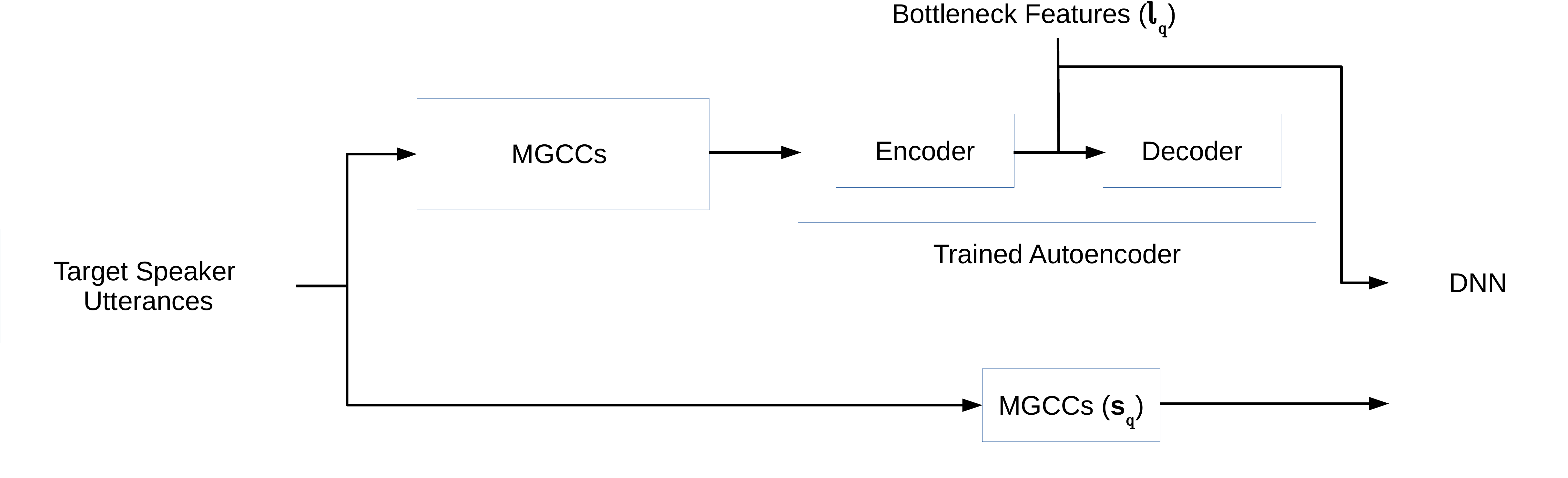}%
				}

	\subfloat[]{%
		  \includegraphics[scale = 0.28]{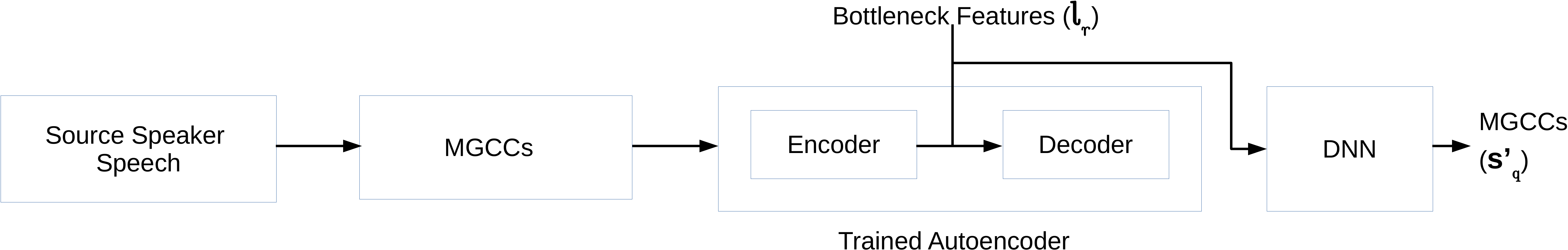}%
				}

	\caption{(a) Training, and (b) Conversion modules in the proposed method.}
\end{figure*}
In this work, we explore the bottleneck features derived from a deep auto-encoder (DAE) as speaker-independent features to perform CLVC. In training phase, a DNN is trained to learn mapping between the target speaker's bottleneck features and spectral features. The bottleneck features are extracted from a pretrained DAE model, trained using multi-speaker data. During conversion, the trained DNN model is used to convert DAE bottleneck features from an arbitrary source speaker to spectral features of the desired target speaker. The proposed approach can capture target speaker-specific characteristics, and does not require any source speaker data to train the VC model. The performance of the proposed method is evaluated using speech data from two Indian languages: Telugu, Tamil and Malayalam. The experimental results show that the proposed method outperform the GMM-based CLVC technique. This paper is organized as follows: Section II gives details of the baseline GMM-based CLVC approach. The proposed CLVC approach is described in Section III. The performance of the proposed method is evaluated and compared with GMM-based CLVC technique in Section IV. Section V summarizes and concludes the present work.

\section{Baseline GMM-based CLVC technique}
For CLVC, approaches which can capture speaker-specific characteristics of a target speaker are required. Such approaches can transform the speech of any arbitary source speaker to a pre-defined target speaker, without recording anything $apriori$ from the source speaker. As mentioned earlier, GMMs can be directly trained with spectral features from target speaker alone to capture speaker-specific charactersitcs. Hence, the GMM-based approach~\cite{b18} is chosen as baseline in this work. The steps in training a GMM to capture target speaker-specific characteristics for CLVC are as follows~\cite{b18}:
\begin{enumerate}
\item Let $L_S$ and $L_T$ denote the source language and target language, respectively.
\item In training phase, extract spectral features ( cepstral coefficients) corresponding to the utterances of $L_T$ and train a GMM with M mixture components that can be used as a tokenizer. The model is denoted as,
\begin{equation}
\lambda _{L_\mathrm{T}} = \{w_{i}, \mu _{i}, {\varSigma }_{i}\};\quad i = 1,2,\ldots ,M 
\end{equation}

here, $w_i$ , $\mu_i$ , and $\Sigma_i$ are the weight, mean vector, and the covariance matrix of the $i_{th}$ mixture component, respectively.
\item During conversion stage, the feature vectors corresponding to the utterances spoken by the source speaker in the source language are extracted. Let $J$ denote the total number of utterances in the $L_S$. From each of these utterances, extract the spectral feature vectors. Let us denote the feature vectors as, $f_k^S$, where $k = 1, 2, ..., N$. Here, N is the total number of feature vectors.
\item For each feature vector of the source speaker, $f_k^S$, given the GMM codebook for the target language $L_T$, the GMM-tokenizer outputs the mean vector ($\mu_k$) of the Gaussian mixture component scoring the highest in GMM likelihood  computation as given below.
\begin{equation}
\mu _{k} = \underset{i=1,2,\ldots ,M}{\arg \max }\left[ w_{i}\cdot g\left( f_{k}^{\mathrm{S}}|\mu _{i},{\varSigma }_{i}\right) \right]
\end{equation}

where,
\begin{equation}
 g\left( f_{k}^{\mathrm{S}}|\mu _{i},{\varSigma }_{i}\right) = \frac{1}{\sqrt{(2\pi )^{D}|{\varSigma }_i|}} e^{-\frac{1}{2} \left( f_{k}^{\mathrm{S}}-\mu _{i}\right) ^{t} {\varSigma }_{i}^{-1} \left( f_{k}^{\mathrm{S}}-\mu _{i}\right) } 
\end{equation}
\item Feature vectors of the source speaker, \(f_{k}^{\mathrm{S}}\), is now replaced by the target feature vectors (codeword), \(\mu _{k}\), with the highest score (likelihood) ensuring the transformation of system  features of the source speaker to that of the target speaker.
\item The fundamental frequency ($F_0$) of the source speaker is transformed to that of the target speaker by a suitable $F_0$ modification factor $F_M$ given by,
\begin{equation}
F_M = \frac{F_T}{F_S}
\end{equation}

where, $F_T$ is the average $F_0$ of target speaker computed from all the training utterances and $F_S$ is the average $F_0$ of the source speaker for a
given utterance. The transformed spectral features and pitch period are given as input to synthesis filter/vocoder for synthesizing the transformed utterance.
\end{enumerate}

\section{Proposed cross-lingual VC Framework}
Training neural network models for capturing target speaker-specific characterisitcs is not as straightforward as with the case of GMMs. The idea in building a neural network model to capture speaker-specific characteristics is as follows~\cite{b4}. Let $l_q$ and $s_q$ be two different representations of the speech signal from a target speaker $q$. While $l_q$ could be interpreted as speaker independent representation of speech signal, $s_q$ could be interpreted as carrying message and speaker information. A mapping function $\Omega(l_q)$ has to be built for transforming $l_q$ to $s_q$. Such a function would be specific to the speaker and could be considered as capturing the essential speaker-specific characteristics. The choice of representation of $l_q$ and $s_q$ is important in developing the mapping functions.
\begin{figure}[]
\centering
	\includegraphics[scale = 0.35]{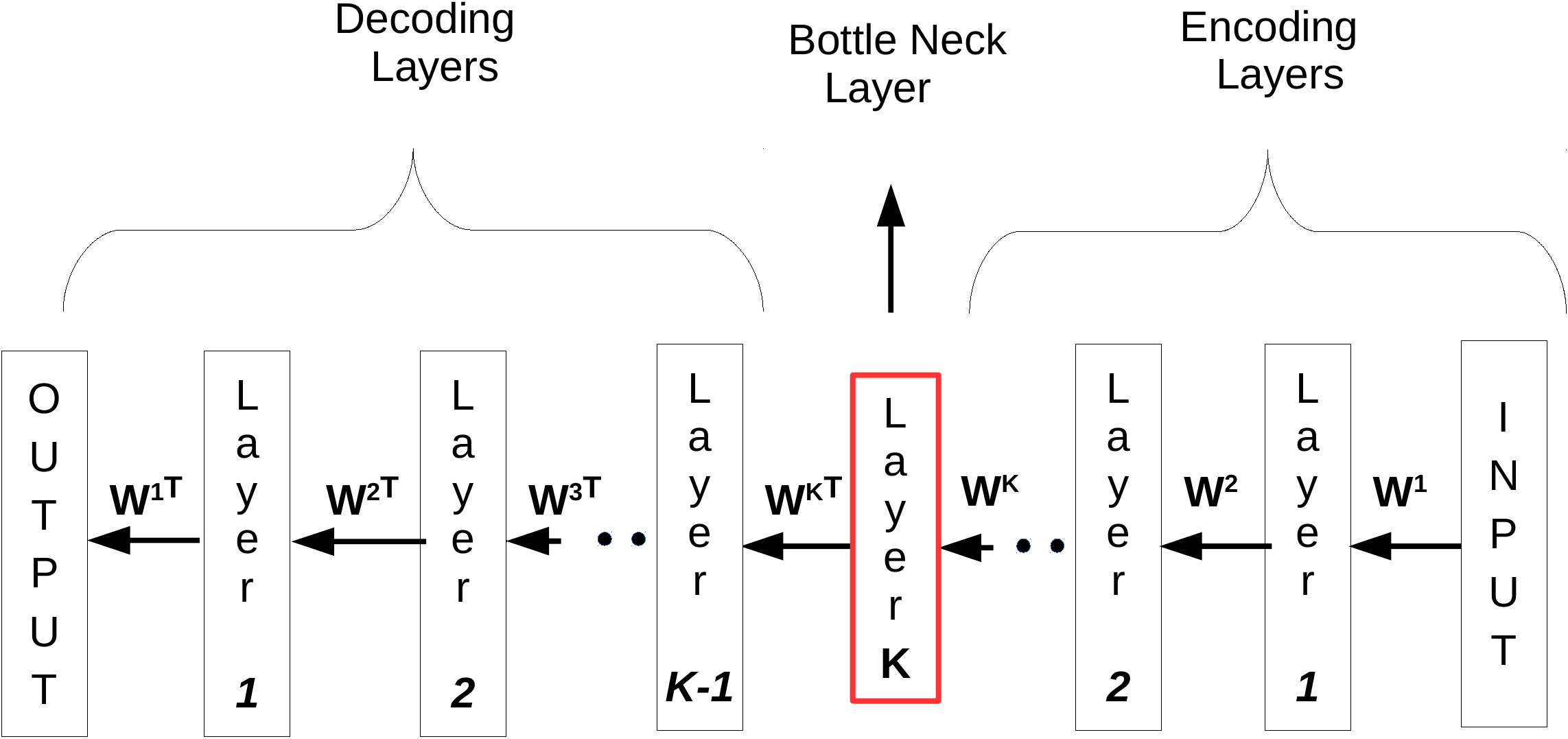}%
	\caption{DAE architecture.}
\end{figure}

Fig. 1 shows the block diagrams of the training and conversion modules of the proposed approach. In the proposed method, $l_q$ and $s_q$ are represented by deep auto-encoder (DAE) bottleneck features and Mel-generalized cepstral coefficients (MGCCs), respectively. The encoder part can cultivate the ability of speaker-independent encoding when an auto-encoder is trained using spectral frames from multiple speakers~\cite{Cc:Hs}. As a result, the encoder can convert an observed frame into latent-variable (or bottleneck features in auto-encoder terminology) which contains information that is irrelevant to speaker, such as phonetic variations. Hence, the bottleneck features have been considered to represent speaker-independent information in the proposed method. In training phase, first bottleneck features corresponding to MGCCs of the target speaker are extracted from a pretrained DAE. The DAE is a feedforward neural network which is pre-trained for speaker-independent bottleneck features generation using a multi-speaker corpus (described in Section III-A). Then, a DNN is trained using back propagation algorithm to minimize the error $||s^{'}_{q}-s_q||^2$, where $s^{'}_{q}=\Omega(l_q)$ and $\Omega(\cdot)$ is the mapping or conversion function. During conversion phase, the trained DNN model is employed to convert $l_r$ to $s^{'}_{q}$, where $l_r$ represents the bottleneck features computed from signal signals of any arbitrary source speaker $r$.
\begin{table*}[]
\centering
\caption{Description of the speech database.}
\label{tab:my-table}
\begin{tabular}{|c|c|c|c|c|c|c|}
\hline
Language   & \multicolumn{3}{c|}{Male}                                                                                                                 & \multicolumn{3}{c|}{Female}                                                                                                               \\ \hline
           & \#Speakers & \begin{tabular}[c]{@{}c@{}}\#Train per\\ speaker\end{tabular} & \begin{tabular}[c]{@{}c@{}}\#Test per\\ speaker\end{tabular} & \#Speakers & \begin{tabular}[c]{@{}c@{}}\#Train per\\ speaker\end{tabular} & \begin{tabular}[c]{@{}c@{}}\#Test per\\ speaker\end{tabular} \\ \hline
Telugu-DAE & 3          & 125 & 30                                                           & 3          & 125                                                           & 30                                                           \\ \hline
Telugu-VC  & 1  (TeM)        & 125 & 30                                                           & 1 (TeF)        & 125                                                           & 30                                                           \\ \hline
Tamil      & 1 (TaM)          & 125 & 30                                                           & 1 (TaM)        & 125                                                           & 30                                                           \\ \hline
Malayalam   & 1 (MaM)         & 125                                                           & 30                                                           & 1 (MaF)          & 125                                                           & 30                                                           \\ \hline
\end{tabular}
\end{table*}
\subsection{Deep Auto-encoder Bottleneck Features}
An Autoencoder (AE) is a feed forward neural network (FFNN) used to learn a representation (encoding) for a set of input data~\cite{b21}. It consists of two blocks: Encoder and Decoder. In a simple auto-encoder having one-hidden-layer, the encoder maps a higher dimensional input vector $\boldsymbol{x}$ to a lower dimensional feature vector $\boldsymbol{y}$ as follows:
\begin{equation}
\boldsymbol{y} = \textit{f}_\theta(\boldsymbol{x}) = s(\boldsymbol{Wx} + \boldsymbol{b})
\end{equation}

Here, $\boldsymbol{y}$ is the bottleneck feature vector representation of the input feature vector $\boldsymbol{x}$. $\theta = \{ \boldsymbol{W}, \boldsymbol{b}\}$ is the encoder parameters. $\boldsymbol{W}$ and $\boldsymbol{b}$ are the weight matrix and a bias vector, respectively. $s$ is a non-linear activation function. The decoder reconstructs the input by using the output $\boldsymbol{y}$ given by the encoder as follows:
\begin{equation}
\boldsymbol{z} = \textit{g}_{\theta^{'}}(\boldsymbol{y}) = s(\boldsymbol{W}^{'}\boldsymbol{y} + \boldsymbol{b}^{'})
\end{equation}
Where, $\theta^{'} = \{\boldsymbol{W}^{'}, \boldsymbol{b}^{'}\}$ is the decoder parameters. $s$ is a linear or non-linear activation function. The weight matrix $\boldsymbol{W}^{'}$ is usually constrained to be the transpose of the matrix in the encoder, i.e., $\boldsymbol{W}^{'} = \boldsymbol{W}^\top$. 

The AE parameters $\{\theta, \theta^{'}\}$ are typically optimized using the mean squared error (MSE) criterion. The model parameters are usually estimated using RMSprop algorithm. An AE can be extended to a deeper architecture by stacking up multiple layers of encoders and decoders, which is called deep auto-encoder (DAE)~\cite{b21}. The additional hidden layers enable the AE to learn mathematically more complex patterns in the data. Fig. 2 shows the basic structure of DAE. In the encoding phase, the units at each hidden layer are calculated given its previous layer as
\begin{equation}
\boldsymbol{y^k} = s \left(\boldsymbol{W^k}\boldsymbol{y^{k-1}}+\boldsymbol{b^k}\right)
 \end{equation}

where $\boldsymbol{W^k}$ and $\boldsymbol{b^k}$ are the parameters of the $\boldsymbol{k}$-th encoder layer, and $\boldsymbol{y}^0=\boldsymbol{x}$. In the decoding phase, the hidden layers are calculated as 
 \begin{equation}
\boldsymbol{y^{k-1}} = s \left(\boldsymbol{{W^k}^\top }\boldsymbol{h^{k}}+\boldsymbol{{b^k}^{\prime }}\right)
 \end{equation}
 and
 \begin{equation}
\boldsymbol{z} =\boldsymbol{{W^1}^\top }\boldsymbol{y^{1}}+\boldsymbol{{b^1}^{\prime }}
\end{equation}
where $\boldsymbol{{W^k}^\top}$ and $\boldsymbol{{b^k}^{\prime }}$ are the parameters of the $\boldsymbol{k}$-th decoder layer, $\boldsymbol{{W^1}^\top}$ and $\boldsymbol{{b^1}^{\prime }}$ are the parameters of the last decoder layer. The training criterion of DAE is the same as AE. 

In this work, we have used DAE with (empirically arrived) architecture $512$-$512$-$\textbf{M}/2$-$512$-$512$, where the enoder has $3$ layers with $\{512,512,\textbf{M}/2\}$ units per layer and the decoder has $2$ layers with $\{512,512\}$ units per layer. The bottleneck features correspond to the output of the last encoding layer (bottleneck layer), which typically contains a small number of neurons relative to the size of the other layers. The input features for the DAE are $\textbf{M}$-dimensional MGCCs. The dimension of bottleneck layer is $\textbf{M}/2$ corresponding to half the dimension of MGCCs, and the dimension of output layer is $\textbf{M}$ corresponding to the dimension of input layer. Note that the dynmaic features were not incorporated into the feature set. Sigmoid activation function is used for all the layers except for the last encoding layer, which has linear activation so that the produced bottleneck features could be real-valued. The DAE is trained under minimum MSE criterion using RMSprop optimizer. The learning rate is set to $0.001$. After training, only the encoder part of DAE is used to generate bottleneck features.  

\subsection{DNN for Feature mapping}
In this work, we used deep neural network (DNN) to capture the functional relationship between the $\textbf{M/2}$-dimensional bottleneck (input) features ($l_q$) and the $\textbf{M}$-dimensional MGCC (output) features ($s_q$) of the given target speaker data. The DNN model used in this paper is a four layer FFNN, and the (empirically arrived) final structure of the network is $\textbf{(M/2)}L\,\, 50N\,\, 50N\,\, \textbf{M}L$, where $L$ denotes a linear unit, and $N$ denotes a non-linear unit. The integer value indicates the number of units used in that layer. The non-linear units use sigmoid activation function. Prior to training, the input and output features are normalized to unit variance and zero mean. The weights of the network are adjusted using backpropagation learning algorithm to minimize the MSE for each pair of input-output features. The learning rate is $0.001$, and the number of epochs is $25$. 

After training, a weight matrix is generated that represents the mapping function between input bottleneck features and output MGCCs. During conversion phase, the obtained weight matrix is used to transform bottleneck features ($l_r$) from any arbitary source speaker $r$ to MGCCs ($s_q^{'}$ ) of the desired target speaker (shown in Figure 1(b)). 

\section{Experiments}
\subsection{Speech Corpus and feature extraction}
For experiments we have considered openslr multi-speaker databases from three Indian languages, namely, Telugu, Tamil and Malayalam. The openslr databases are available for free download at \url{https://www.openslr.org/resources.php}. For experiments, we chose 8 speakers (4 male and 4 female speakers) from Telugu language, 2 speakers (1 male and 1 female speakers) from Tamil language, and 2 speakers (1 male and 1 female speakers) from Malayalam language. Each speaker has 125 utterances for training and 30 utterances for testing. Data were recorded at 48 kHz, but we have downsampled to 16 kHz. The details of data considered for training and testing are given in Table I. As shown in table, six out of the eight speakers (3 female and 3 male speakers) from Telugu dataset were considered for training DAE and remaining speakers were used in VC experiments. The reason for choosing this database for DAE training is explained in the following section. 
 
The WORLD vocoder~\cite{b20} was used to exctract speech parameters: $f_0$, aperiodicity (AP), and Spectral Envelope (SE). The frame length was 25 ms and the frame shift was 5 ms. The FFT length was set to 1024, so the resulting SE and AP were both $513$-dimensional. 40-dimensional MGCCs were derived from each spectral envelope. The dynamic features were not appended to the feature set. The proposed and baseline VC models were trained using features corresponding to the target speaker alone. The trained models were then used to map the MGCCs of an arbitary source speaker to the MGCCs of the target speaker. The transformed MGCCs were converted back to $513$-dimensional SE. The $f_0$ of the source speaker was converted to that of the target speaker as in the baseline system (described in section II). The AP of source speaker was kept unmodified. Finally, all speech parameters were given as input to WORLD vocoder to synthesize the transformed utterance.

\subsection{Training baseline and proposed CLVC systems}
In the baseline system, a GMM having 128 components is trained using 40-dimensional MGCCs from target speaker alone. The speaker-independent features are not required for a GMM-based system. The baseline system transforms voice of an arbitary source speaker by replacing the source feature vectors with the mean vector of the Gaussian mixture component scoring the highest in GMM likelihood. In this work, a GMM model is built for every speaker from each language.

The training procedure of proposed VC system is different from that of the baseline system. Prior to training VC model, first a DAE is trained using speech utterances from 6 Telugu speakers to learn speaker-independent representations. The regional Indian languages considered are acoustically similar, to certain extent~\cite{Br:Ca}. Hence, a common phoneset is derived by exploiting the acoustic similarities across the Indian languages~\cite{Br:Ca}. In total, there are 39 phonemes in Tamil, 48 in Telugu, and 48 in Malayalam. It is observed that, 37 phones are common to all the languages. This intutively shows that a DAE trained with data from one of these languages will be good enough to generate speaker-independent features in case of all the languages. Comparing Telugu and Malayalam, there are 47 phones in common. Comparing, Telugu and Tamil, there are 37 phones in common and 11 are unique to Telugu and 2 to Tamil. Similarly, comparing Malayalam and Tamil, there are 38 phones in common and 10 are unique to Malayalam and 1 is unique to Tamil. Considering the unique and common phones, either Malayalam or Telugu language  is a better choice than Tamil language, for training DAE. Hence, we chose Telugu speaker data for training the DAE model. The encoder receives MGCCs computed from all the speakers and converts them to bottleneck features. The decoder reconstructs the input from the bottleneck features. The training procedure is terminated when there is no further improvement in terms of MSE for 15 epochs. The trained encoder is then utilized to extract bottleneck features for voice conversion. A DNN model which maps bottleneck features to MGCCs of a target speaker is built separately for all the considered speakers, except for those used during DAE training.

\subsection{Performance Evaluation}
Performance of the CLVC systems is evaluated using two subjective measures, namely, ABX preference test and Comparative Mean Opinion Score (MOS). In preference tests, subjects were asked to listen to a pair of converted speech utterances and chose the one that is closer to natural speech in terms of similarity in voice. 
\begin{figure}[]
\centering
	\includegraphics[width=0.7\columnwidth]{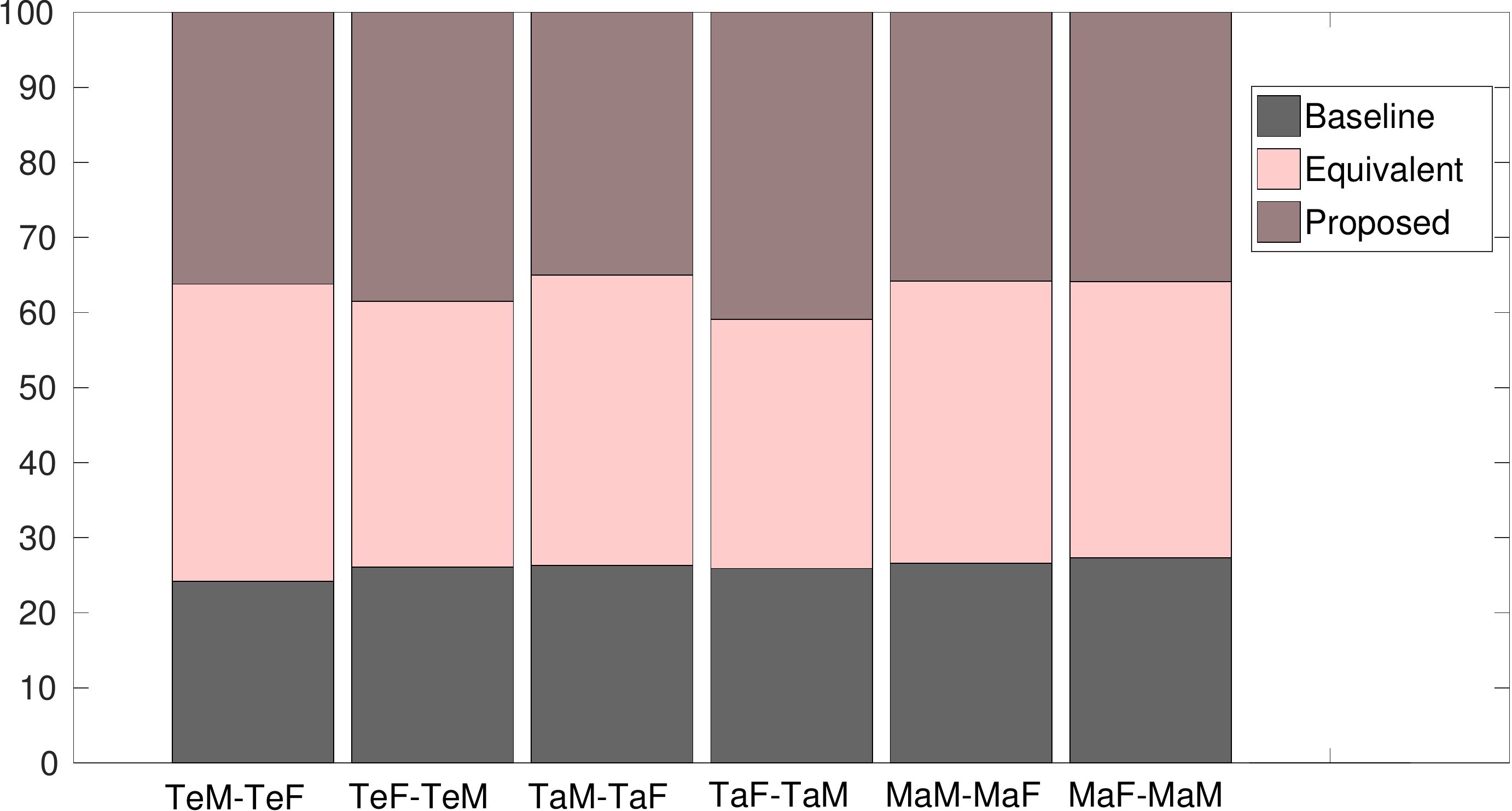}%
	\caption{Preference test results (\%) for speaker similarity on intra-lingual voice conversion.}
\end{figure}
 \begin{figure}[]
\centering
	\includegraphics[width=0.7\columnwidth]{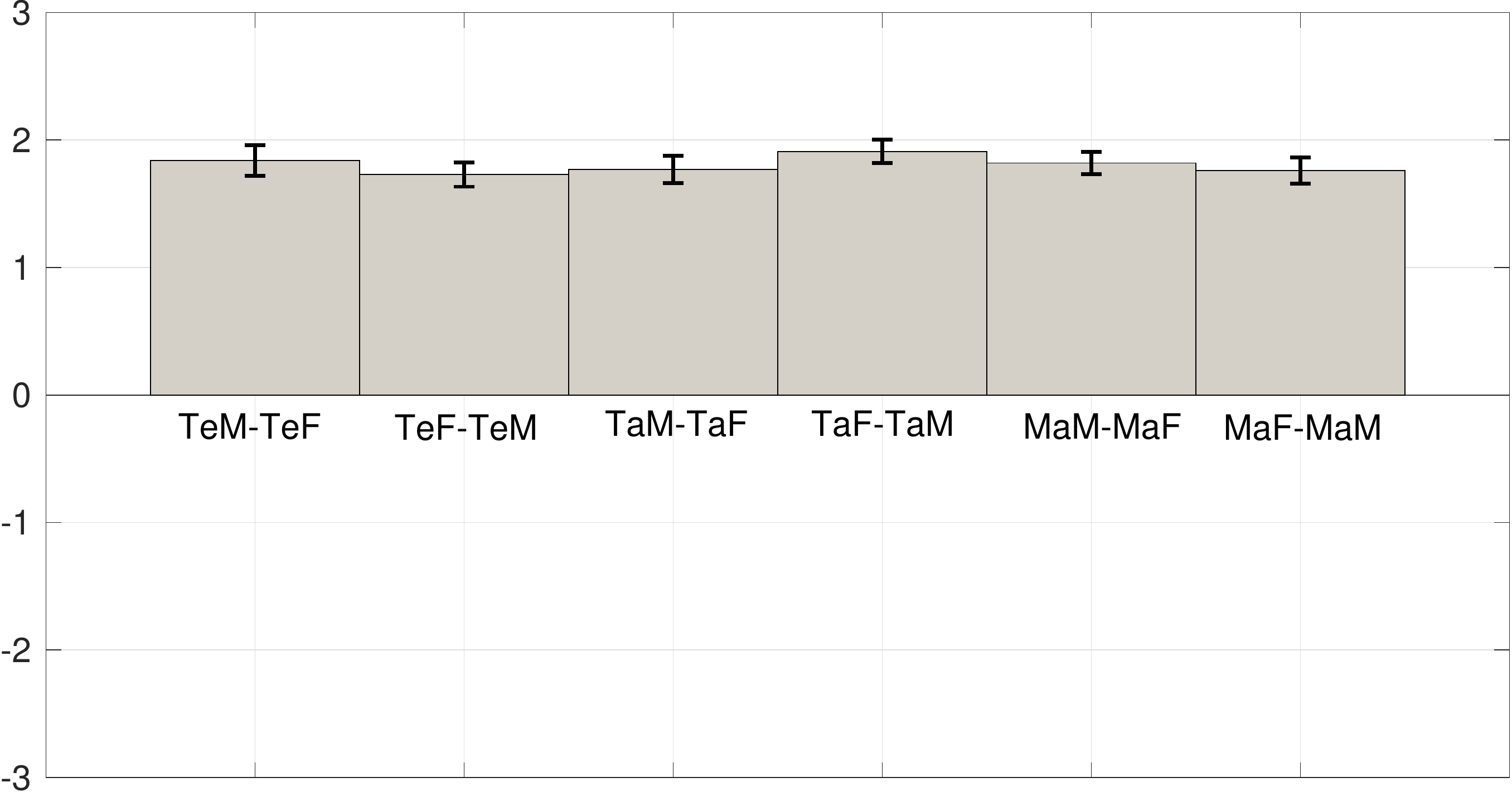}%
	\caption{CMOS scores with 95\% confidence intervals obtained on intra-lingual voice conversion.}
\end{figure}
\begin{figure}[]
\centering
	\includegraphics[width=0.7\columnwidth]{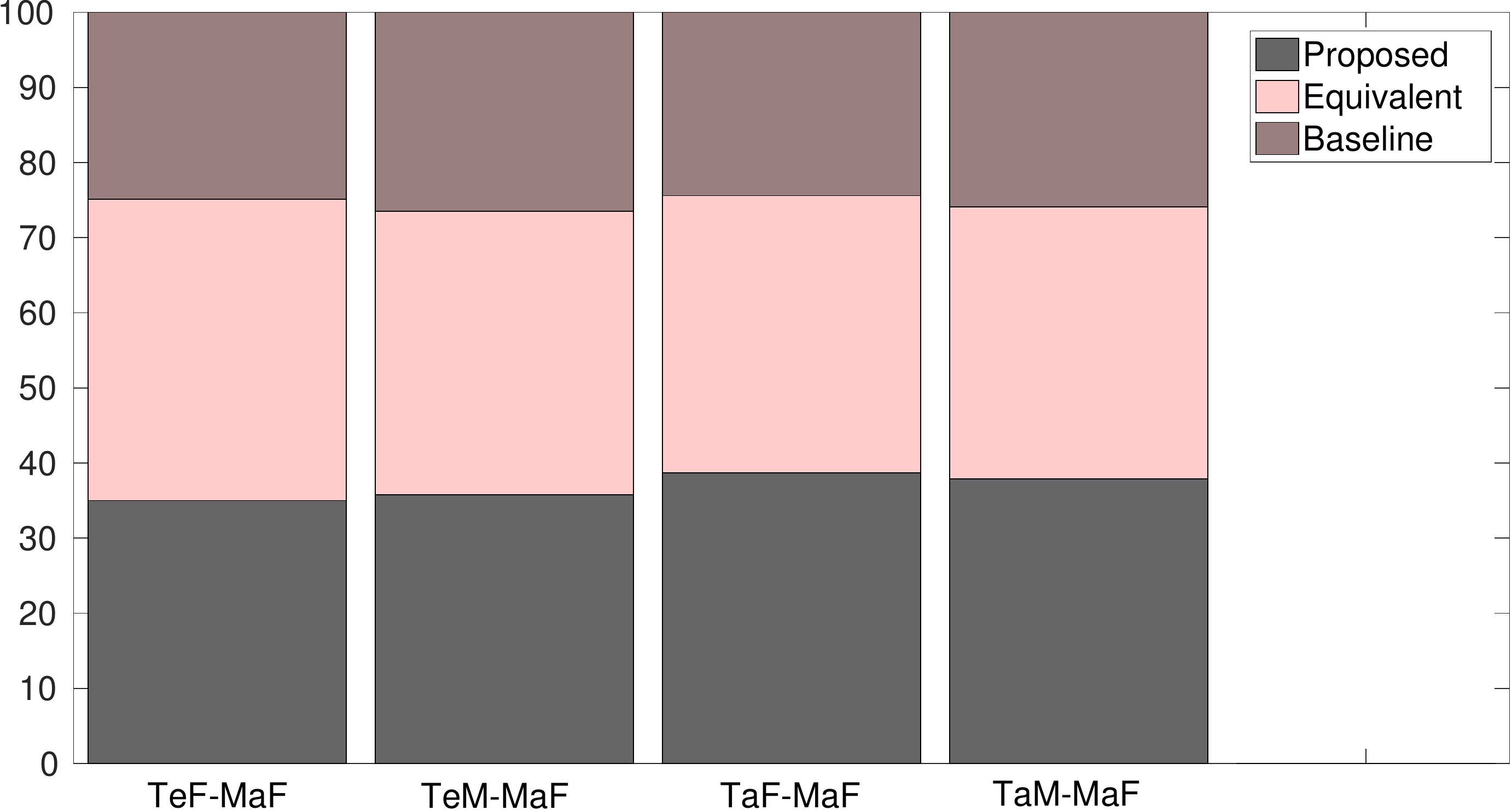}%
	\caption{Preference test results (\%) for speaker similarity on cross-lingual voice conversion.}
\end{figure}
 \begin{figure}[]
\centering
	\includegraphics[width=.7\columnwidth]{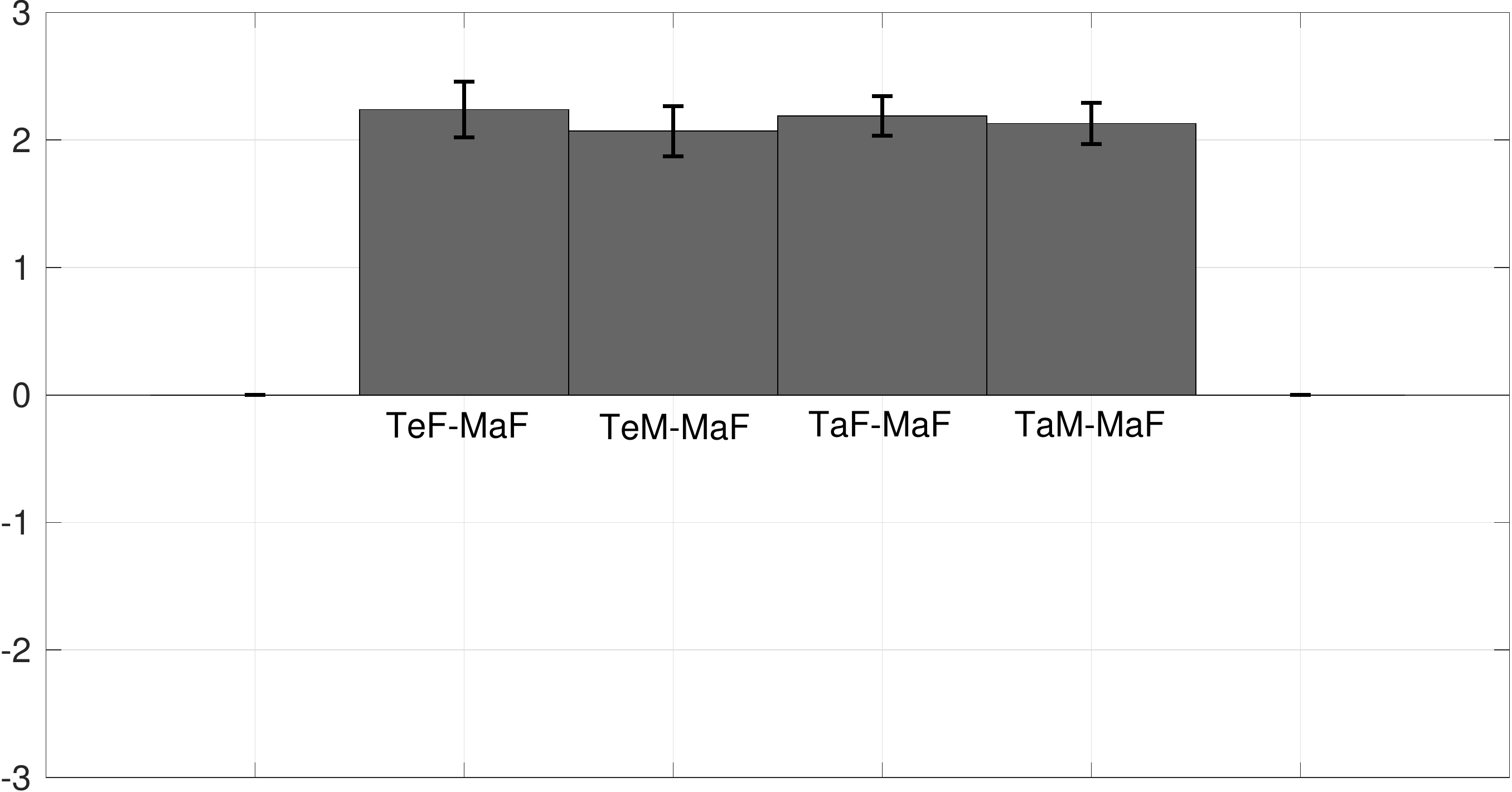}%
	\caption{CMOS scores with 95\% confidence intervals obtained on cross-lingual voice conversion.}
\end{figure}
In CMOS test, subjects have to listen to two converted speech utterances (one from the proposed approach and another from one of the baseline method, randomly shuffled) and then they have to rate the difference between the two samples on a 7-point scale ranging from much worse (-3) to much better (3). Subjects preference for proposed method over the baseline method and the opposite is indicated by a positive CMOS score and negative CMOS score, respectively. 15 listeners, including 5 Telugu, 5 Tamil and 5 Malayalam, participated in all the tests. The listening tests were conducted in the laboratory environment by playing the speech signals through headphones.

The results of ABX preference test and CMOS test for intra-lingual VC and cross-lingual VC are provided in Figures 4 \& 5 and Figures 6 \& 7, respectively. To reduce burden on listeners, we have considered only Malayalam Female as target and speakers from remaining languages as source, in cross-lingual VC. That is, the malayalam female speaker can now speak Tamil and Telugu. The trends in the results can be analyzed as follows: (1) The CMOS scores show that the proposed method provides a better voice quality compared to the baseline on intralingual VC, with a significance level of $p < 0.001$. The CMOS scores are further improved on cross-lingual VC. (2) Although a trend seems to favor proposed method compared to baseline in speaker similarity, the difference is not significantly different. The listeners reported that there is muffledness and more distortion in the speech converted with baseline GMM-based system.
The improved perceptual quality with proposed method is mainly due to better acoustic feature mapping. The results also indicate that the proposed approach can render high-quality speaker conversion irrespective of languages. Hence, our system is capable of addressing the issues associated with unaligned data sets. 

\section{Summary and conclusion}
In this paper, we have presented a DAE-based cross-lingual voice conversion approach that can capture target speaker specific characteristics. The DAE is trained with data from multiple speakers to learn speaker-independent representations. Even though data from only one language is used to train DAE, the encoder performs robustly across acoustically closer languages. To build VC model for a given target speaker, first the MGCCs are passed through the encoder to derive bottleneck features. Then, a DNN is trained to predict MGCCs of target speaker from bottleneck features. The proposed approach does not require data from source speaker, and can map spectral features of any arbitary source speaker onto a target speaker's acoustic space. Hence, the proposed method can be considered as a ``many-to-one mapping" method. The performance of the CLVC systems is evaluated using three acoustically similar Indian languages. The results of subjective evaluation confirm that both quality and target speaker similiarity of converted speech from proposed CLVC system are much better than the baseline GMM-based CLVC system. In future, we plan to utilize the proposed CLVC technique to develop a polyglot SPSS system for Indian languages.

 \balance




%

\end{document}